\documentstyle[preprint,aps]{revtex}
\renewcommand{\theequation}{\thesection.\arabic{equation}}
\begin{document}
\title{Weyl's Character Formula for $SU(3)$ - A Generating Function Approach}
\author{J. S. Prakash{\thanks {jsp@iopb.ernet.in}}} 
\address{Institute Of Physics\\ Sachivalaya Marg, Bhubaneswar 751 005, India}
\date{March 1996}
\maketitle

\begin{abstract} 
Using a generating function for the Wigner's $D$-matrix elements of
$SU(3)$ Weyl's character formula for $SU(3)$ is derived using
Schwinger's technique.
\end{abstract}
PACS number(s): 

\section{Introduction}
In a previous paper we had set up a calculus to do compuatations
on the group $SU(3)$.  This is a calculus which is similar to
the one developed for $SU(2)$ by Schwinger\cite{SJ},
Bargmann\cite{BV}.  With the help of this calculus we were able
to solve some important problems in group theory related to
$SU(3)$.  The basic aim of this paper is to apply this calculus
to obtain Weyl's character formula for the group $SU(3)$ and
thus show that, using our calculus, doing computations on
$SU(3)$ is as easy as dealing with $SU(2)$.  

A formula for the characters of unitary groups was derived by
Weyl\cite{WH} using integration on group manifolds.  It was also
obtained by Freudenthal\cite{JN} using a purely Lie algebraic
method.  Boerner\cite{BH} derives the same result by examining
Young Frames.  (See also the books of Littlewood and Hamermesh
(\cite{DEL,MH})).  This formula is the starting point for
computing the dimensions of the Irreducible Representations
(IRs) and also for arriving at the branching theorems for the
IRs(\cite{HB,GM}).  In this paper Weyl's formula, for the group
$SU(3)$, is derived, for the first time, from a generating
function for the matrix elements of $SU(3)$.  The generating
function method was successfully applied by Schwinger\cite{SJ}
for the computation of Weyl's character formula for the group
$SU(2)$ long back.  In essence we are extending his method to
$SU(3)$.

The plan of the paper is as follows.  In section 1 we collect
some results from our previous work on the IRs of $SU(3)$.  The
next section, that is section 2, is devoted to the computation
of the normalizations of the basis states.  We then derive
Weyl's character formula for $SU(3)$ using Schwinger's method in
section 3.  Section 4 is devoted to a discussion of the results
of this paper.  We reproduce Schwinger's original derivation of
Weyl's formula for the characters of $SU(2)$ in the appendix.

\section{Overview of our previous results}

In this section we briefly review the results that we need on
the group $SU(3)$.  Some of these results were obtained by us in
a previous paper {\cite{JSPHSS}}.

$SU(3)$ is the group of $3\times 3$ unitary unimodular matrices
$A$ with complex coefficients. It is a group of $8$ real
parameters.  The matrix elements satisfy the following
conditions

\begin{eqnarray}
A&=&(a_{ij}), \nonumber \\ A^\dagger A &=& I, \qquad AA^\dagger
= I\, ,\,\,\, \mbox{where}\, I\, \mbox{is the identity matrix}\nonumber \\
\mbox{det}(A)&=&1\, .
\label{A}
\end{eqnarray}

\subsection{Parametrization}

One well known parametrization of $SU(3)$ is due to
Murnaghan {\cite{mfd}}, see also {\cite{bmarh,cemm,ntj,bakd}}.
In this we write a typical element of $SU(3)$ as :

\begin{equation}
D(\delta_1,\delta_2,\phi_3) U_{23}(\phi_2,\sigma_3) U_{12}(\theta_1,\sigma_2) 
U_{13}(\phi_1,\sigma_1)\, ,
\end{equation}
with the condition $\phi_3 = -(\delta_1 + \delta_2)$.  Here $D$
is a diagonal matrix whose elements are ${\exp}(i\delta_1)$,
${\exp}(i\delta_2)$ ,${\exp}(i\phi_3)$ and $U_{pq}(\phi,\sigma)$ is a $3
\times 3$ unitary unimodular matrix which for instance in the
case $p=1$, $q=2$ has the form
\begin{equation}
\pmatrix{{\cos}\phi               & -{\sin}\phi {\exp}(-i\sigma) & 0 \cr
         {\sin} \phi {\exp}(i\sigma) & {\cos} \phi               & 0 \cr
         0                     & 0                      & 1}
\end{equation}
The $3$ parameters $\phi_1$, $\phi_2$, $\phi_3$ are longitudinal
angles whose range is $-\pi \leq \phi_i \leq \pi$, and the
remaining $6$ parametrs are latitude angles whose range is
$\-\frac{1}{2} \pi \leq \sigma_i \leq \frac{1}{2}\pi$.

Now the trnasformations $U_{23}$ and $U_{13}$ can be changed into
transformations of the type $U_{12}$ whose matrix elements are
known, by the following device
\begin{eqnarray}
U_{13}(\phi_1, \sigma_1) &=& (2,3) U_{12}(\phi_1,\sigma_1) (2,3)\, , \nonumber \\
U_{23}(\phi_2, \sigma_3) &=& (1,2) (2,3) U_{12}(\phi_2,\sigma_3) (2,3) (1,2)\, ,
\end{eqnarray}
where $(1,2)$ and $(2,3)$ are the transposition matrices
\begin{equation}
(1,2) = \pmatrix{0 & 1 & 0 \cr
                 1 & 0 & 0 \cr
                 0 & 0 & 1 }, \hspace{.1in} (2,3) = \pmatrix{0 & 1 & 0 \cr
                                                      1 & 0 & 0 \cr
                                                      0 & 0 & 1 }\, .
\end{equation}

In this way the expression for an element of the $SU(3)$ group becomes
\begin{equation}
D(\delta_1,\delta_2,\phi_3) (1,2) (2,3) U_{12}(\phi_2,\sigma_3)
(2,3) (1,2) U_{12}(\theta_1,\sigma_2) (2,3)
U_{12}(\phi_1,\sigma_1) (2,3)\, .
\end{equation}

\subsection{Lie algebra Generators}

Later in order to calculate the normalizations of our
un-normalized basis states, we have to first to obtain the
representation of the generators of the Lie algebra of $SU(3)$
as differential operators.  For this purpose and also for
parametrizing the equivalence classes of $SU(3)$ we take our
$SU(3)$ generators to be

\begin{eqnarray}
\pi^0 =\pmatrix {1 & 0 & 0 \cr 0 & -1 & 0\cr 0 & 0 & 0},\,\,\, 
\eta^0=\pmatrix {1 & 0 & 0 \cr 0 & 1 & 0\cr 0 & 0 & -2},\,\,\,  
\pi^- =\pmatrix {0 & 1 & 0 \cr 0 & 0 & 0\cr 0 & 0 & 0},\,\,\,
\pi^+=\pmatrix {0 & 0 & 0 \cr 1 & 0 & 0\cr 0 & 0 & 0},\,\,\, \nonumber\\
\nonumber\\
K^-  =\pmatrix {0 & 0 & 1 \cr 0 & 0 & 0\cr 0 & 0 & 0},\,\,\, 
K^+  =\pmatrix {0 & 0 & 0 \cr 0 & 0 & 0\cr 1 & 0 & 0},\,\,\, 
K^0  =\pmatrix {0 & 0 & 0 \cr 0 & 0 & 1\cr 0 & 0 & 0},\,\,\, 
{\bar K}^0=\pmatrix {0 & 0 & 0 \cr 0 & 0 & 0\cr 0 & 1 & 0}\, .\nonumber\\
\label{IGRs}
\end{eqnarray}

\subsection{Irreducible Representations.}

Our parametrization provides us with a defining
irreducible representation $\underline{3}$ of $SU(3)$ acting on
a $3$ dimensional complex vector space spanned by the triplet
$z_1, z_2, z_3$ of complex variables.  The hermitian adjoint of
the above matrix gives us another defining but inequivalent
irreducible representaion $\underline{3^*}$ of $SU(3)$ acting on
the triplet ${w_1,w_2,w_3}$ of complex variables spanning
another $3$ dimensional complex vector space.  Tensors
constructed out of these two $3$ dimensional representations
span an infinite dimensional complex vector space.  \\

\subsection{{{The Constraint}}}
If we impose the constraint

\begin{eqnarray}
 z_1w_1+z_2w_2+z_3w_3=0\, ,
\label{z.w}
\end{eqnarray}
on this space we obtain an infinite dimensional complex vector
space in which each irreducible representation of $SU(3)$ occurs
once and only once.  Such a space is called a model space for
$SU(3)$.  Further if we solve the constraint
$z_1w_1+z_2w_2+z_3w_3=0$ and eliminate one of the variables, say
$w_3$, in terms of the other five variables $z_1, z_2, z_3, w_1,
w_2$ we can write a genarating function to generate all the
basis states of all the IRs of $SU(3)$.  This generating
function is computationally a very convenient realization of the
basis of the model space of $SU(3)$.  Moreover we can define a
scalar product on this space by choosing one of the variables,
say $z_3$, to be a planar rotor ${\exp}(i\theta)$.  Thus the
model space for $SU(3)$ is now a Hilbert space with this
('auxiliary') scalar product between the basis states.  The
above construction was carried out in detail in a previous paper
by us {\cite{JSPHSS}}.  For easy accessability we give a
self-contained summary of some results which are relevant for us
here.

\subsection{Labels for the basis states.}

\noindent {{\bf{(i). Gelfand-Zetlein labels}}}

Normalized basis vectors are denoted by,
$\vert{M,N;P,Q,R,S,U,V}>$.  All labels are non-negative
integers.  All Irreducible Represenatations(IRs) are uniquely
labeled by $(M, N)$.  For a given IR $(M, N)$, labels
$(P,Q,R,S,U,V)$ take all non-negative integral values subject to
the constraints:

\begin{equation}
R+U=M\hspace{.1in},\hspace{.1in} S+V=N\hspace{.1in},\hspace{.1in} P+Q=R+S.
\label{GZL}
\end{equation}

The allowed values can be presribed easily: $R$ takes all values
from $0$ to $M$, and $S$ from $0$ to $N$.  For a given $R$ and
$S$, $Q$ takes all values from $0$ to $R+S$.

Gelfand-Zetlein basis states are represented by the triangular
patterns or arrays\\

\begin{eqnarray}
\pmatrix{m_{13} & & m_{23} & & 0 \cr
         & m_{12} & & m_{22} \cr
          & & m_{11}}
\label{GZpat}
\end{eqnarray}     
\\
The entries are non-negative integers satisfying the betweenness
constraints 

\begin{eqnarray}
(i) m_{13} \geq m_{12} \qquad (ii) m_{12} &\geq& m_{23} \qquad
(iii) m_{23} \geq m_{22} \nonumber 
\\
(iv) m_{22} \geq 0 \qquad (v) m_{12} &\geq& m_{11} \qquad
(vi) m_{11} \geq m_{22} 
\label{betw}
\end{eqnarray}

In terms of these labels our labels $P, Q, R, S, U, V$ can be
expressed as

\begin{eqnarray}
M &=& m_{13} - m_{23}\, ,\nonumber\\ N &=& m_{23} \,
,\nonumber\\ P &=& \frac{1}{2}(m_{11} + m_{23}) - \frac{1}{2}
\vert m_{11} - m_{23} \vert - m_{22} + \vert m_{11} - m_{23} \vert\epsilon(m_{11}-m_{23})\, ,\nonumber\\
Q & = & m_{12} - \frac{1}{2}(m_{11} + m_{23}) - \frac{1}{2}
\vert m_{11} - m_{23} \vert + \vert m_{11} - m_{23} \vert\epsilon(-(m_{11}-m_{23}))\, ,\nonumber\\
P + Q &=& m_{12} - m_{22}\, ,\nonumber\\
R     &=& M - m_{22}     \, ,\nonumber\\
S     &=& m_{12} - M     \, ,\nonumber\\
U     &=& m_{22}         \, ,\nonumber\\
V     &=& m_{13} - m_{12}\, ,\nonumber\\
\label{our-GZ}
\end{eqnarray}

\noindent {{\bf{(ii). Young - Weyl Tableu labels}}}.

The Gelfand-Zetlein basis may also be represnted by the Young -
Weyl tableu with two rows of boxes labelled by 1's, 2's, and 3's
as follws 

\begin{eqnarray}
&\#& {\mbox{ of}}\cdots         \qquad 1{\mbox{st row}}\quad\qquad 2{\mbox{nd row}}\nonumber\\
&\#& {\mbox{ of }} 1'{\mbox{s}} \qquad m_{11}\quad\qquad          \qquad {\mbox{zero}}    \, ,\nonumber\\
&\#& {\mbox{ of }} 2'{\mbox{s}} \qquad m_{12} - m_{11} \qquad m_{22}  \, ,\nonumber\\
&\#& {\mbox{ of }} 3'{\mbox{s}} \qquad m_{13} - m_{12} \qquad m_{23} - m_{22}
\label{YWT}
\end{eqnarray}

\noindent {{\bf{(iii). Quark model labels}}}.

The relation between the above Gelfand-Zetlein labels and the
Quark Model labels is as given below.

\begin{eqnarray}
2I  &=&P+Q=R+S\, ,\nonumber\\
2I_3&=&P-Q    \, ,\nonumber\\
Y   &=&\frac{1}{3} (M-N) + V-U\, , \nonumber\\
&&= \frac{2}{3} (N-M)-(S-R)\, .
\label{QML}
\end{eqnarray}

\subsection{Explicit realization of the basis states}
 
\noindent {\bf{(i). {Generating function for the basis states of $SU(3)$}}}

The generating function for the basis states of the IR's of
$SU(3)$ can be written as
 
\begin{equation}
g(p,q,r,s,u,v)={\exp}(r(pz_1+qz_2)+s(pw_2-qw_1)+uz_3+vw_3)\, .
\label{g()}
\end{equation}

The coefficient of the monomial $p^Pq^Qr^Rs^Su^Uv^V$ in the
Taylor expansion of Eq.(\ref{g()}), after eliminating $w_3$
using Eq.(\ref{z.w}), in terms of these monomials gives the
basis state of $SU(3)$ labelled by the quantum numbers $P, Q, R,
S, U, V$.

\noindent {\bf{{ (ii). Formal generating function for the basis
states of $SU(3)$ }}}

The generating function Eq.(\ref{g()}) can be written formally
as 

\begin{equation}
g=\sum_{P,Q,R,S,U,V} p^Pq^Qr^Rs^Su^Uv^V \vert PQRSUV)\, ,
\label{FGF}
\end{equation}
where $\vert PQRSUV)$ is an unnormalized basis state of $SU(3)$
labelled by the quantum numbers $P,Q,R,S,T,U,V$.

Note that the constraint $P+Q=R+S$ is automatically satisfied in
the formal as well as explicit Taylor expansion of the generating
function.

\noindent {\bf{{ (iii). Generalized generating function for the basis
states of $SU(3)$}}}

It is useful, while computing the normalizations(see below) of the basis
states, to write the above generating function in the following
form

\begin{equation}
{\cal G}(p,q,r,s,u,v)={\exp}(r_pz_1+r_qz_2+s_pw_2+s_qw_1+uz_3+vw_3)\, .
\label{GGF}
\end{equation}

In the above generalized generating function (\ref{GGF}) the
following notation holds.

\begin{equation}
r_p=rp, \qquad r_q=rq, \qquad s_p=sp, \qquad s_q=-sq\, .
\label{rprqspsq}
\end{equation}

\subsection{Notation}
Hereafter, for simplicity in notation we assume all variables
other than the $z^i_j$ and $w^i_j$ where $i, j=1,,2,3$ are real
eventhough at some places we have treated them as complex
variables.  Our results are valid even without this restriction
as we are interested only in the coefficients of the monomials
in these real variables rather than in the monomials themselves.

\subsection{'Auxiliary' scalar product for the basis states.}

The scalar product to be defined in this section is 'auxiliary'
in the sense that it does not give us the 'true' normalizations
of the basis astes of $SU(3)$.  However it is compuataionally
very convenient for us as all compuattaions with this scalar
product get reduced to simple Gaussian integrations and the
'true' normalizations themselves can then be got quite easily.  

{\bf{(i). Scalar product between generating functions of basis
states of $SU(3)$}}

We define the scalar product between any two basis states in
terms of the scalar product between the corresponding generating
functions as follows :

\begin{eqnarray}
(g', g)&=& {\int_{-\pi}^{+\pi}}{\frac{d\theta}{2\pi}} \int
\frac{d^{2}z_1}{\pi^2} \frac{d^{2}z_2}{\pi^2}
\frac{d^{2}w_1}{\pi^2} \frac{d^{2}w_2}{\pi^2}
{\exp}(-\bar{z_1}z_1 - \bar{z_2}z_2 - \bar{w_1}w_1
-\bar{w_2}w_2)\nonumber\\
&&\nonumber \\
&&\times {\exp}((r'(p'z_1+q'z_2) + s'(p'w_2-q'w_1) - \frac{-v'}{z_3}
(z_1w_1 + z_2w_2) + u'\bar{z}_3) \nonumber \\
\nonumber\\
&&\times {\exp}((r(pz_1 + qz_2) + s(pw_2-qw_1) - \frac{-v}{z_3}
(z_1w_1 + z_2w_2) + uz_3)\, , \nonumber \\
&&\nonumber \\
&&= (1-v'v)^{-2} \left (\sum_{n=0}^{\infty}
\frac{(u'u)^n}{(n!)^2}\right )
{\exp}\left [(1-v'v)^{-1}(p'p + q'q)(r'r + s's)\right ]\, .
\label{gg1}
\end{eqnarray}

{\bf{(ii). Modifed scalar product}}

In order to facilitate comparison with Schwinger's computation,
of the Weyl's character formula for $SU(2)$, and to be able to
see, a little more clearly, the possibilities of generalizations
to higher groups we redefine the $\theta$ part of the scalar
product  and write the above Eq.(\ref{gg1}) as

\begin{eqnarray}
(g', g)&=& (1-v'v)^{-2} {\exp}\left [ \frac{(p'p + q'q)(r'r + s's)+u^{'}u}{(1-v'v)}\right ]\, .
\label{gg'}
\end{eqnarray}

{\bf{(iii). Choice of the variable $z_3$}}

To obtain the Eqs.(\ref{gg1}, \ref{gg'}) we have made the choice
\begin{eqnarray}
z_3=\exp(i\theta )\, .
\label{z3}
\end{eqnarray}

The choice, Eq.(\ref{z3}), makes our basis states for $SU(3)$
depened on the variables $z_1,z_2,w_1,w_2$ and $\theta $.

{\bf{(iv). Scalar product between the gneralized generating
functions of the basis states of $SU(3)$}} 

For the generalized generating function the scalar product
becomes 

\begin{eqnarray}
{(\cal{G'}, \cal{G})} &=& (1 - v'v)^{-2} {\exp}\left [(1 - v'v)^{-1}({r_p}'r_p +
{r_q}'r_q + {s_p}'s_p + {s_q}'s_q) \right ]\nonumber \\
&&\nonumber \\
&&\times \left [{\sum_{n=0}}^{\infty} \frac{1}{(n!)^2}\left (u'- v
\frac{({r_p}'{s_q}'+ {r_q}'{s_p}')}{(1 - v'v)}\right )^n\, {\bf \cdot}\,\left (u - v'
\frac{({r_p}{s_q} + {r_q}{s_p})}{(1 - v'v)}\right )^n \right ]\,
,
\label{GIP1}
\end{eqnarray}
and as in Eq.(\ref{rprqspsq})

\begin{eqnarray}
r_p=rp, \qquad r_q=rq, \qquad s_p=sp, \qquad s_q=-sq\, ,\nonumber \\
r_p'=r'p', \qquad r_q'=r'q', \qquad s_p'=s'p', \qquad
s_q'=-s'q'\, .
\label{s_q=-sq}
\end{eqnarray}

{\bf{(v). Modified scalar product between the gneralized generating
functions of the basis states of $SU(3)$}}

With our modified scalar product Eq.(\ref{GIP1}) reads as 

\begin{eqnarray}
{(\cal{G'}, \cal{G"})} 
&=& (1 - v'v")^{-2} {\exp} 
\left [ 
\frac{({r_p}'r_p" +{r_q}'r_q" + {s_p}'s_p" + {s_q}'s_q")}{(1 -
v'v")} \right. \nonumber \\ 
&&\nonumber \\ 
&&+ \left.  \left ( u'- v"\frac{({r_p}'{s_q}' + {r_q}'{s_p}')}{(1 - v'v")}\right ) 
            \left (u" - v'\frac{({r_p}"{s_q}" + {r_q}"{s_p}")}{(1 - v'v")}\right ) 
\right ]\, .
\label{GIP}
\end{eqnarray}

The notation Eq.(\ref{s_q=-sq}) holds good in this equation
also but only for the singly primed and the unprimed variables
(hidden in doubly primed ones).


\section{Computaion of the Normalizations}

The noramalizations using the scalar product Eq.(\ref{gg1}) were
computed by us in our paper\cite{JSPHSS}.  But since we have now
changed, the $\theta$ part of the, scalar product we have to
compute these normalizations once again.  This is done, as
before, by the requirement that the representation matrix be
unitary in each irreducible representation.

Now let

\begin{eqnarray}
E&=&{\mbox {The set of quantum numbers used in the basis}}\, ,\nonumber \\
\vert E )&=&{\mbox {Unnormalized basis state}}\, ,\nonumber \\
\vert E >&=&{\mbox {Normalized basis state}}\, ,\nonumber \\
(E'\Vert E)&=&M(E)\delta_{E'E}\, ,\nonumber\\ 
&=&{\mbox{Scalar product between unnormalized basis states with respect}}\, \nonumber \\
&&{\mbox{to the auxiliary scalar product}}\, ,\nonumber \\ 
T&=&{\mbox {Generator of SU(3)}}\nonumber\, , \\ 
(E'\Vert T\vert E)&=&{\mbox {Matrix element of}}\,\, T\,\, {\mbox {between
unnormalized basis states}}\,\, E',\,\, E\,\, \nonumber \\
&&{\mbox {with respect to the 'auxiliary' scalar product}}\, .
\end{eqnarray}

The symbol $\Vert$ stands for the scalar product with
respect to the 'auxiliary' measure.

Then assuming that 

\begin{eqnarray}
\vert E )&=&N^{\frac{1}{2}}(E) \vert E >\, ,
\end{eqnarray}
it can be shown that,
\begin{eqnarray}
\left \vert \frac{N(E)}{N(E')}\right \vert 
&=& \frac{(E'\Vert T\vert E)M(E)}
{(E\Vert T^*\vert E')^*M(E')}\, ,
\label{NbyN}
\end{eqnarray}
where $T^*$ is the adjoint of $T$ and $(\Vert )^*$ is the
complex conjugate of $(\Vert )$.

Thus we can fix normalizations using an
'auxiliary' scalar product which allows explicit computations even
though it is not the 'true' scalar product.  

{\bf {(i). Generators of $SU(3)$.}}

The 'auxiliary' matrix elements of $SU(3)$ are to be computed by
the action of the generators of $SU(3)$ on the generating
function of the unnormalized basis states.  For this purpose the
generators of $SU(3)$ are needed.  The generators of $SU(3)$
take a particularly convenient form when realized in terms of
differential operators in the variables $r_p, r_q, u, s_p, s_q,
v$.  These operators act on the unnormalized basis states
generated by the generalized generating function\, ${\cal
{G}}$\,in Eq.({\ref{GGF}}).  These generators are listed below.

\begin{eqnarray}
{\hat \pi}^0&=&r_p\frac{\partial }{\partial
r_p}-r_q\frac{\partial }{\partial r_q}-s_q\frac{\partial
}{\partial s_q}+s_p\frac{\partial }{\partial s_p}\, ,\nonumber \\
{\hat \pi}^-&=&r_p\frac{\partial }{\partial
r_q}-s_p\frac{\partial }{\partial s_q}\, ,\nonumber \\
{\hat \pi}^+&=&r_q\frac{\partial }{\partial
r_p}-s_q\frac{\partial}{\partial s_p}\, ,\nonumber \\
{\hat K}^-&=&r_p\frac{\partial }{\partial
u}-v\frac{\partial}{\partial s_q}\, ,\nonumber \\
{\hat K}^+&=&u\frac{\partial }{\partial
r_p}-s_q\frac{\partial}{\partial v}\, ,\nonumber \\
{\hat K}^0&=&r_q\frac{\partial }{\partial
u}-v\frac{\partial}{\partial s_p}\, ,\nonumber \\
{\hat {\bar {K}}}^0&=&u\frac{\partial }{\partial
r_q}-s_p\frac{\partial}{\partial v}\, ,\nonumber \\
{\hat \eta}&=&r_p\frac{\partial }{\partial
r_p}+r_q\frac{\partial }{\partial r_q}+s_q\frac{\partial
}{\partial s_q}+s_p\frac{\partial }{\partial
s_p}-2u\frac{\partial }{\partial u}+2v\frac{\partial }{\partial
v}\, .
\label{Genrs}
\end{eqnarray}

{\bf {(i). 'Auxiliary' Matrix Elemets of Generators of $SU(3)$.}}\\

Consider ${\hat \pi}^- $ as given in Eq.(\ref{Genrs}).  We have

\begin{eqnarray}
\left (g',\,\, {\hat \pi}^-g\right ) &=&p{\bar q}'\frac{({\bar r}'r
+{\bar s}'s)}{(1-{\bar v}'v)}\left (g',\,\,g\right )\, .
\end{eqnarray}

This gives us 

\begin{eqnarray}
(P,Q+1,R,S,U,V\Vert {\hat \pi}^-\vert P+1,
Q,,R,S,U,V)&=&M_3(P,Q,R,S,U,V)\nonumber \\ {\mbox
{Similarly~~~~~~~~~~~~~~~~~~~~~~~~~~~~~~~~~~~~~~~~~~~~~~}}&&\nonumber \\ 
(P+1,Q,R,S,U,V\Vert {\hat \pi}^+\vert P,Q+1,R,S,U,V)
&=&M_3(P,Q,R,S,U,V)\nonumber \\\, .
\end{eqnarray}

We now compute the relative normalizations implied by ${\hat
K}^{\pm}$.  As we said above, to calculate, $(g',\,\,{\hat
K}^-g)$ we use the generalized generating function
\begin{eqnarray}
(g',\,\,{\hat K}^-g)&=&(r_p\frac{\partial }{\partial
u}-v\frac{\partial }{\partial s_q})({\cal
{G}'},\,\,{\cal{G}}){\bf \vert}\, ,
\end{eqnarray}
where the vertical line at the end of this equation means that
after applying differential operator on
$({\cal{G}'},\,\,{\cal{G}})$, we need to set the values
Eq.(\ref{s_q=-sq}).  For instance,

\begin{eqnarray}
(r_ps_q+r_qs_p){\bf \vert}&=&0\, ,\nonumber \\
({\bar r}_p'{\bar s}_q'+{\bar r_q}'{\bar s_p}'){\bf \vert}&=&0\, .
\end{eqnarray}

We get

\begin{eqnarray}
(g',\,\,{\hat K}^-g)&=&
\frac{1}{(1 - v'{\bar v})^2}
{\exp}\left ( {\frac{(pp'+qq')(rr'+ss')}{1-v'{\bar v}}} + {\bar u}'u
\right )\nonumber \\
&&\times \left (rp{\bar u}' + \frac{rp{\bar u}'v{\bar
v}'}{(1-{\bar v}'v}+v\frac{{\bar s}'{\bar q}'}{(1-{\bar v}'v}\right )\, .
\end{eqnarray}

Let 

\begin{eqnarray}
M_1(,P,Q,R,S,U,V) &=& {\mbox {coefficient of }}
{p'}^P{q'}^q{r'}^R{s'}^S{u'}^U{v'}^V \times p^Pq^Qr^Rs^Su^Uv^V\nonumber \\
&&{\mbox {in the expansion of }} \frac{\left(g',\,\, g\right
)}{(1-{\bar v}'v)}\,.\nonumber \\
\end{eqnarray}

Assume similar meanings to the coefficients 
$M, M_2, M_4, M_5, {\mbox{ and }} M_6$. (See Table below.)

\begin{center}
\begin{tabular}{|c|c|c|}\hline
$M(\cdots )$ & $\frac{\exp[(1-v'{\bar
v})^{-1}(p'p+q'q)(r'r+s's)+u'u]} {(1-v'{\bar v})^2}$ &
$\frac{(2I+1+V)!} {P!Q!R!S!U!V!(2I+1)}$ \\ \hline $M_1(\cdots )$
& $\frac{\exp[(1-v'{\bar v})^{-1}(p'p+q'q)(r'r+s's)+u'u]}
{(1-v'{\bar v})^3}$ & $\frac{(2I+2+V)!}
{P!Q!R!S!U!V!(2I+1)(2I+2)}$\nonumber \\ \hline 
$M_2(\cdots )$ & $\frac{\exp[(1-v'{\bar v})^{-1}(p'p+q'q)(r'r+s's)+u'u]}
{(1-v'{\bar v})^4}$ & $\frac{(2I+3+V)!}
{P!Q!R!S!U!V!(2I+1)(2I+2)(2I+3)}$\nonumber \\ \hline 
$M_3(\cdots )$ & $(rr'+ss')\times\frac{\exp[(1-v'{\bar
v})^{-1}(p'p+q'q)(r'r+s's)+u'u]} {(1-v'{\bar v})^3} $ & {\mbox
Not needed.}
\nonumber  \\ \hline 
$M_4(\cdots )$ & $  (pp'+qq')(rr'+ss') \times\frac{\exp[(1-v'{\bar v})^{-1}(p'p+q'q)(r'r+s's)+u'u]]}
{(1-v'{\bar v})^4}$ 
& $\frac{(2I+2+V)!}{P!Q!R!S!U!V!}\frac{2I}{(2I+1)(2I+2)}$\nonumber  \\ \hline 
$M_5(\cdots )$ & $\frac{\exp[(1-v'{\bar v})^{-1}(p'p+q'q)(r'r+s's)]}
{(1-v'{\bar v})^2}$ 
& $\frac{(2I+1+V)!}{P!Q!R!S!U!V!}\frac{1}{(2I+1)(2I+2)}$\nonumber  \\ \hline  
$M_6(\cdots )$ & $\frac{\exp[(1-v'{\bar v})^{-1}(p'p+q'q)(r'r+s's)+u'u]}
{(1-v'{\bar v})^3}$ 
& $\frac{(2I+2+V)!}{P!Q!R!S!U!V!}\frac{1}{(2I+1)(2I+2)}$\nonumber  \\ \hline  
\end{tabular}
\end{center}
where in the above table in the first column $(\cdots )=(P,Q,R,S,U,V)$.

Matching coefficients of like powers we get (see Table above.),

\begin{eqnarray}
(P,Q+1,R,S+1,U,V\Vert{\bar K}^-\vert
P,,Q,R,S,U,V+1)&=&M_1(,P,Q,R,S,U,V)\, ,\nonumber \\
(P,Q,R,S,U+1,V)\Vert{\hat K}^-\vert
P+1,Q,R+1,S,U,V)&=&M_5(P,Q,R,S,U,V)\nonumber \\
&&+M_6(P,Q,R,S,U,V-1)\, .
\label{KM}
\end{eqnarray}

Similarly,

\begin{eqnarray}
(P+1,Q,R+1,S,U+1,V)\Vert{\hat K}^+\vert
P,Q,R,S,U+1,V)&=&M_1(P,Q,R,S,U,V)\, ,\nonumber \\
(P,Q,R,S,U,V+1\Vert{\hat K}^+\vert
P,Q+1,R,S+1,U,V)&=&M_6(,P,Q,R,S,U-1,V)\nonumber \\
&&+2M_1(P,Q,R,S,U,V)\nonumber \\
&&+M_4(P,Q,R,S,U,V)\, .
\label{KP}
\end{eqnarray}

Likewise

\begin{eqnarray}
(P,Q,R,S,U+1,V)\Vert{\hat K}^0\vert
P,Q+1,R+1,S,U,V)&=&M_5(P,Q,R,S,U,V)\nonumber \\
&&+M_6(,P,Q,R,S,U,V-1)\, ,\nonumber \\
(P+1,Q,R,S+1,U,V\Vert{\bar K}^0\vert
P,Q,R,S,U,V+1)&=&-M_1(,P,Q,R,S,U,V)\, .\nonumber \\
\label{KZ}
\end{eqnarray}
and
\begin{eqnarray}
(P,Q+1,R+1,S,U,V)\Vert{\hat {\bar K}}^0\vert
P,Q,R,S,U+1,V)&=&M_1(P,Q,R,S,U,V)\, ,\nonumber \\
(P,Q,R,S,U,V+1\Vert{\hat {\bar K}}^0\vert
P+1,Q,R,S+1,U,V)&=&-M_6(,P,Q,R,S,U-1,V)\nonumber \\
&&-2M_1(P,Q,R,S,U,V)\nonumber \\
&&-M_4(P,Q,,RS,,U,V)\, .
\label{KZB}
\end{eqnarray}

\subsection{New Normalizations}

In this subsection we calculate the normalizations of
unnorrmalized basis states by requiring that the matrix elements
of the generators of $SU(3)$ be hermitian adjoints with respect
to each other in pairs.  

Accordingly, by using Eq.(\ref{NbyN}) together with Eq.(\ref{KM}) and
Eq.(\ref{KP}) we get
\begin{eqnarray}
\left \vert \frac{N(P+1,Q,R+1,S,U,V)}{N(P,,Q,R,S,U+1,V)}\right
\vert &=&\frac{(V+2I+2)(U+1)(2I+1)}{(P+1)(R+1)(2I+2)}\label{R1}\\
&&\nonumber \\
\left \vert \frac{N(P,Q+1,R,S+1,U,V)}{N(,P,Q,R,S,U,V+1)}\right
\vert&=& \frac{(U+2I+2)(V+1)(2I+1)}{(Q+1)(S+1)(2I+2)}\label{R2}\, ,
\end{eqnarray}
and by applying Eq.(\ref{NbyN}, this time, to Eq.(\ref{KZ}) and
Eq.(\ref{KZB}) we get,
\begin{eqnarray}
\left \vert \frac{N(P,Q+1,R+1,S,U,V)}{N(P,,Q,R,S,U+1,V)}\right
\vert &=&   \frac{(V+2I+2)(U+1)(2I+1)}{(Q+1)(R+1)(2I+2)} \label{R3}\\
&&\nonumber \\
\left \vert \frac{N(P+1,Q,R,S+1,U,V)}{N(,P,Q,R,S,U,V+1)}\right
\vert&=& \frac{(U+2I+2)(V+1)(2I+1)}{(P+1)(S+1)(2I+2)}\label{R4}\, .
\end{eqnarray}

The normalization constant $N(P,Q,R,S,U,V))$ is uniquley fixed
by the above constraints Eqs.(\ref{R1}-\ref{R4}).  The solution
is given by 
\begin{eqnarray}
N(P,Q,R,S,U,V)&=&\frac{(U+2I+1)!(V+2I+1)!}{P!Q!R!S!U!V!(2I+1)}\, .
\label{N}
\end{eqnarray}

From the above Eq.(\ref{N}) we see that the normalization for
the basis states is the same as that obtained in our erlier
work\cite{JSPHSS} though our present inner product is slighly
different from the one used in that reference. 

Below we briefly summarize the results of this section.

\noindent {\bf{{(i). 'Auxiliary' normalizations of unnormalized
basis states}}} 

The scalar product between two unnormalized basis states,
computed using our 'auxiliary scalar product, is given by

\begin{eqnarray}
M(PQRSUV)&\equiv &(PQRSUV\vert PQRSUV)\, ,\nonumber \\
&&=\frac{(V+P+Q+1)! }{P! Q! R!S! U! V! (P+Q+1)}\, .
\label{M}
\end{eqnarray}

\noindent {\bf{(ii). Scalar product between the unnormalized and normalized
basis states}}

The scalar product, computed using our 'auxiliary' scalar
product, between an unnormalized basis state and a normalized
one is denoted by $(PQRSUV\Vert PQRSUV>$ and is given below

\begin{equation}
(PQRSUV\Vert PQRSUV>=N^{-1/2}(PQRSUV)\times M(PQRSUV)\, .
\label{(||)}
\end{equation}

\noindent {\bf{{(iii). 'True' normalizations of the basis
states}}} 

We call the ratio of the 'auxiliary' norm of the unnormalized
basis sate represented by $\vert PQRSUV)$ and the scalar product
of the normalized Gelfand-Zeitlin state, represented by $\vert
PQRSUV > $, with the unnormalized basis state as 'true'
normalization.  It is given by

\begin{eqnarray}
N^{1/2}(PQRSUV)&\equiv & \frac{(PQRSUV\Vert PQRSUV)}{(PQRSUV\Vert PQRSUV>}\, ,\nonumber \\
&&=(\frac{(U+P+Q+1)! (V+P+Q+1)! }{P! Q! R!S! U! V! (P+Q+1)})^{1/2}\, .
\end{eqnarray}

Before leaving this section we just note that the new
normalizations computed by using our slighly different scalar
product for the basis states are not much different from the
ones computed using our previous scalar product.  

\section{{{Generating function for the Wigner's $D$-matrix 
elements of $SU(3)$.}}}

In this section we first derive, very briefly, a generating
function for the Wigner's $D$-matrix elements of $SU(3)$.  This
generating function will be our starting point in the next
section.

We know from Eq.(\ref{FGF}) that 

\begin{equation}
g(p,q,r,s,u,v,z_1,z_2,w_1,w_2) = \sum_{P,Q,R,S,U,V}
p^Pq^Qr^Rs^Su^Uv^V\vert PQRSUV)\, ,
\end{equation}
where $\vert PQRSUV)$ is an unnormalized basis state in the IR
labeled by the two positive integers $(M=R+U, N=S+V)$.

We know from Eq.(\ref{N}),

\begin{equation}
\vert PQRSUV) = N^{(1/2)}(PQRSUV) \vert PQRSUV>\, ,
\end{equation}

where $2I=P+Q$ and $\vert PQRSUV)$ is a normalized basis state.

Therefore

\begin{equation}
g=\sum_{PQRSUV} \left (
\frac{(U+2I+1)! (V+2I+1)! } {P!Q!R!S!U!V!(2I+1)} \right )^{(1/2)} p^Pq^Qr^Rs^Su^Uv^V \vert PQRSUV>\, .
\end{equation}

Now,

\begin{eqnarray}
Ag(p,q,...)&=& \sum_{PQRSUV}\sum_{P'Q'R'S'U'V'}\left 
( \frac{(U+2I+1)!(V+2I+1)!}{P!Q!R!S!U!V!(2I+1)}\right )^{(1/2)}\nonumber \\
&& \nonumber \\
&&\times D_{PQRSUV,\,\, P'Q'R'S'U'V'}^{(M=R+U,\,\, N=S+V)}\times \,\,
p^Pq^Qr^Rs^Su^Uv^V \times \,\,\vert PQRSUV>\, .
\end{eqnarray}

To get a generating function for the matrix elements alone we
have to take the inner product of this transformed generating
function with the generating function for the basis states.


Thus,

\begin{eqnarray}
&&\left ( g(p", q", r", s", u", v";\,\, z_1, z_2, z_3, w_1, w_2),\,\,
A g(p, q, r, s, u, v;\,\, z_1, z_2, z_3, w_1, w_2)\right )\nonumber \\
&& \nonumber \\
&=& \sum_{PQRSUV} \sum_{P'Q'R'S'U'V'}
\sum_{P"Q"R"S"U"V"} \left
(\frac{(U+2I+1)!(V+2I+1)!}{P!Q!R!S!U!V!(2I+1)}\right
)^{(1/2)}\nonumber \\ 
&& \nonumber \\
&& \times  (P"Q"R"S"U"V"\Vert P'Q'R'S'U'V'>\times 
D_{PQRSUV,\,\, P'Q'R'S'U'V'}^{(M=R+U,\,\,N=S+V)}(A)\nonumber \\
&& \nonumber \\
&& \times p^Pq^Qr^Rs^Su^Uv^V p"^{P"}q"^{Q"}r"^{R"}s"^{S"}u"^{U"}v"^{V"}\, .
\end{eqnarray}

But we know from Eq.(\ref{(||)}),

\begin{eqnarray}
& (&P"Q"R"S"U"V" \Vert P'Q'R'S'U'V'>\nonumber \\
&& \nonumber \\
& = & \left (\frac{(U' + 2I' + 1)! (V' + 2I' + 1)!}{P'! Q'! R'! S'! U'! V'! (2I' + 1)}\right )^{(-1/2)}
\times \frac{(V'+P'+Q'+1)!}{P'!Q'!R'!S'!U'!V'!(P'+Q')}\nonumber \\
&& \nonumber \\
&&\times \delta_{P"P'} \delta_{Q"Q'} \delta_{R"R'} \delta_{S"S'}
\delta_{U"U'} \delta_{V"V'}\, .
\end{eqnarray}

Substituting this formula and changing the double primed
variables to single primed ones, we get

\begin{eqnarray}
&&\left ( g(p', q', r,' s', u', v';\,\, z_1,z_2,z_3,w_1,w_2){\bf ,}\quad 
 Ag(p,q,r,s,u,v;\,\, z_1,z_2,z_3,w_1,w_2) \right ) \nonumber \\
&& \nonumber \\
& = & \sum_{{\tiny {PQRSUV;\,\,P'Q'R'S'U'V'}}} \left
(\frac{(U+2I+1)!(V+2I+1)!} {P! Q! R! S! U! V! (2I +1)})\times
 (\frac{P'! Q'! R'! S'! U'! V'! (2I' + 1)}{(U'+2I'+1)!(V'+2I'+1)!}\right )^{(1/2)}
\nonumber \\
&&\nonumber \\
&&\times (\frac{(V'+P'+Q'+1)!}{P'!Q'!R'!S'!U'!V'!(P'+Q'+1)})
\times D_{PQRSUV,\,\,P'Q'R'S'U'V'}^{(M=R+U,\,\,N=S+V)}(A) \nonumber \\
&&\nonumber \\
&& \times p^P q^Q r^R s^S u^U v^V{p'}^{P'}{q'}^{Q'}{r'}^{R'}{s'}^{S'}{u'}^{U'}{v'}^{V'}\, .
\end{eqnarray}

Next we calculate this inner product using the explicit
realization for the generating function.  For this purpose it is
advantageous, as will be seen in a minute, to use the
generalized generating function for the basis states

\begin{eqnarray}
{\cal G} &=& {\exp}(r_pz_1+r_qz_2+s_pw_2+s_qw_1+uz_3+vw_3)\nonumber\\ 
&& \nonumber \\
& =& {\exp}\left ( {\pmatrix{r_p & r_q & u}}{\pmatrix{z_1\cr z_2 \cr z_3}} +
{\pmatrix{w_1 & w_2 & w_3}}{\pmatrix{s_q \cr s_p \cr v}}\right
)\, .
\end{eqnarray}

When any element $A \in SU(3)$ acts on this generating function it undergoes
the following transformation

\begin{eqnarray}
A{\cal G} = {\exp}\left ( {\pmatrix{r_p & r_q & u}} A{\pmatrix{z_1\cr z_2 \cr z_3}} +
{\pmatrix{w_1 & w_2 & w_3}}A^{\dagger} {\pmatrix{s_q \cr s_p \cr
v}}\right )\, .
\end{eqnarray}

As is clear from the above equation we can let the triplets
$r_p, r_q, u$ and $s_q,s_p,v$ undergo the transformation instead
of the triplets $z_1,z_2,z_3$ and $w_1,w_2,w_3$.  Therfore we
can write the transformed generating function as

\begin{eqnarray}
A{\cal G} = {\cal G}(r_p",r_q",u"; s_q",s_p",v")\, ,
\end{eqnarray}

where

\begin{eqnarray}
r_p"&=& a_{11}r_p+a_{21}r_q+a_{31}u\nonumber \\
r_q"&=& a_{12}r_p+a_{22}r_q+a_{32}u\nonumber \\
u"&=& a_{13}r_p+a_{23}r_q+a_{33}u\, ,\nonumber \\
&&  \nonumber \\
s_q"&=& a^*_{11}s_q+a^*_{21}s_p+a^*_{31}v\nonumber \\
s_p"&=& a^*_{12}s_q+a^*_{22}s_p+a^*_{32}v\nonumber \\
v"&=& a^*_{13}s_q+a^*_{23}s_p+a^*_{33}v\, .
\label{A*GGF}
\end{eqnarray}



To continue with our computation we have to take the inner
product of this transformed generating function with the
(untransformed) generating function of the basis states.

This is known to us from Eq.(\ref{GIP}) as

\begin{eqnarray}
{(\cal{G'}, \cal{G"})} 
&=& (1 - v'v")^{-2} {\exp} 
\left [ 
\frac{({r_p}'r_p" +{r_q}'r_q" + {s_p}'s_p" + {s_q}'s_q")}{(1 -
v'v")} \right. \nonumber \\ 
&&\nonumber \\ 
&&+ \left.  \left ( u'- v"\frac{({r_p}'{s_q}' + {r_q}'{s_p}')}{(1 - v'v")}\right ) 
            \left (u" - v'\frac{({r_p}"{s_q}" + {r_q}"{s_p}")}{(1 - v'v")}\right ) 
\right ]\, .
\end{eqnarray}

This expression gets further simplified if we substitute from
Eq.(\ref{rprqspsq})

\begin{eqnarray*}
r_p'=r'p', \quad r_q'=r'q', \quad s_q'=-s'q', \quad s_p'=s'p\, .
\end{eqnarray*}

We, therefore, get

\begin{eqnarray}
{(\cal{G'}, \cal{G"})} &=& (1 - v'v")^{-2} {\exp} \left [ (1 - v'v")^{-1}({r_p}'r_p" +
{r_q}'r_q" + {s_p}'s_p" + {s_q}'s_q") \right. \nonumber \\
&& \nonumber \\
&& + \left.  u'(u" - v'\frac{({r_p}"{s_q}" + {r_q}"{s_p}")}{(1 - v'v")}) \right ]\, .
\label{GG'}
\end{eqnarray}

One last simplification can be brought about in the above expression
when we recognize that

\begin{eqnarray}
 {r_p}"{s_q}" + {r_q}"{s_p}"+u"v"  = & {r_p}{s_q} + {r_q}{s_p}+vu\, ,\nonumber\\
 = & vu\, .
\end{eqnarray}

This tells us that

\begin{eqnarray}
 {r_p}"{s_q}" + {r_q}"{s_p}"= uv - u"v"\, .
\end{eqnarray}

Substituting this in our expression Eq.(\ref{GG'}) for the inner
product we get,

\begin{eqnarray}
{(\cal{G'}, \cal{G"})} &=& (1 - v'v")^{-2} {\exp} \left [ (1 -
v'v")^{-1}({r_p}'r_p" + {r_q}'r_q" + {s_p}'s_p" + {s_q}'s_q")
\right. \nonumber \\
&& \nonumber \\
&& + \left.  u'(u" - v'\frac{({u}{v} - {u}"{v}")}{(1 - v'v")}) \right ]\, ,\nonumber \\
&& \nonumber \\
& =& (1 - v'v")^{-2} {\exp} \left [ (1 - v'v")^{-1}({r_p}'r_p" +
{r_q}'r_q" + {s_p}'s_p" + {s_q}'s_q") \right. \nonumber \\ 
&& \nonumber \\
&& + \left. \left ( {u'} \frac{({u}" - {u}{v}v')}{(1 - v'v")} \right )\right ]\, .
\label{GFWD1}
\end{eqnarray}

The expression on the right hand side of Eq.(\ref{GFWD1}) is our
generating function for the Wigner's $D$-matrix elements of
$SU(3)$.

Hereafter we denote the generating function Eq.(\ref{GFWD1}) for
the Wigner's $D$-matrix elements of $SU(3)$ by the symbol ${\cal
G}(D(A))$. 

\section{Weyl's Character Formula for $SU(3)$}

We now apply Schwinger's method (see appendix), for the case of
$SU(2)$, to obtain Weyl's formula for the characters of IRs of
$SU(3)$.

Weyl's character formula for (\cite{WH,HB,GM}) $U(3)$ is written
as
\begin{eqnarray}
\chi^{(m_{13}, m_{23}, m_{33})}_{U(3)} ((D(A))) = \frac{
\left \vert
\pmatrix
{
  \epsilon^{m_{13}+2}_1 & \epsilon^{m_{23}+1}_1   & \epsilon^{m_{33}}_1 \cr 
  \epsilon^{m_{13}+2}_2 & \epsilon^{m_{23}+1}_2   & \epsilon^{m_{33}}_2 \cr
  \epsilon^{m_{13}+2}_3 & \epsilon^{m_{23}+1}_3   & \epsilon^{m_{33}}_3
}
\right \vert
}{
\left \vert
\pmatrix
{
\epsilon^{2}_1 & \epsilon_1 & 1 \cr 
\epsilon^{2}_2 & \epsilon_2 & 1 \cr 
\epsilon^{2}_3 & \epsilon_3 & 1
}
\right \vert
}
\, ,
\label{wchi}
\end{eqnarray}
where $\epsilon_k= = \exp(-i\theta_{kk})$.  

From Eq.(\ref{wchi}) it can be shown\cite{GM} that

\begin{eqnarray}
\chi^{(m_{13}-m_{23}, m_{23}-m_{33},0)}_{SU(3)} (D(A)) &=& e^{i\theta_{33}(m_{13}+m_{23}+m_{33})} 
\sum^{m_{13}}_{m^{'}_{12}=m_{23}} \sum^{m_{23}}_{m^{'}_{22}=m_{33}} 
e^{-3i\theta_{33}(\frac{m^{'}_{12}+m^{'}_{22}}{2})} \nonumber\\
\nonumber\\
&&\times \frac{\sin \frac{m^{'}_{12}-m^{'}_{22}+1}{2}(\theta_{11} - \theta_{22})}
{\sin \frac{1}{2}(\theta_{11}-\theta_{22})}\, .\nonumber\\
\label{gmchi}
\end{eqnarray}
where the representations of $SU(3)$ are given by 
\begin{eqnarray}
(p, q, 0) = (m_{12}-m_{23}, m_{23}-m_{33},0)\, .
\end{eqnarray}

Below we will try to obtain the form represented by
Eq.(\ref{gmchi}) for Weyl's character formula for $SU(3)$.

Our starting for applying Schwinger's method is the generating
function for the Wigner's $D$-matrix elements of $SU(3)$
Eq.(\ref{GFWD1}) and its formal equivalent Eq.(\ref{FEQ}).
These two forms are reproduced below.

\begin{eqnarray}
{\cal G}(D(A))& = & \sum_{{\tiny {PQRSUV;\,\,P'Q'R'S'U'V'}}} \left
(\frac{(U+2I+1)!(V+2I+1)!} {P! Q! R! S! U! V! (2I +1)})\times
 (\frac{P'! Q'! R'! S'! U'! V'! (2I' + 1)}{(U'+2I'+1)!(V'+2I'+1)!}\right )^{(1/2)}
\nonumber \\
&&\nonumber \\
&&\times (\frac{(V'+P'+Q'+1)!}{P'!Q'!R'!S'!U'!V'!(P'+Q'+1)})
\times D_{PQRSUV,\,\,P'Q'R'S'U'V'}^{(M=R+U,\,\,N=S+V)}(A) \nonumber\\
\nonumber\\
&&\times p^P q^Q r^R s^S u^U v^V{p'}^{P'}{q'}^{Q'}{r'}^{R'}{s'}^{S'}{u'}^{U'}{v'}^{V'}\, , \nonumber\\
\\
\label{FEQ} 
&=& (1 - v'v")^{-2} {\exp} \left [ 
\frac{({r_p}'r_p" +{r_q}'r_q" + {s_p}'s_p" + {s_q}'s_q")
+u'(u" - uvv')}{(1 - v'v")}\right ]\, .\nonumber\\
\end{eqnarray}

We first take up the formal expression, r.h.s. of
Eq.(\ref{FEQ}).  Since $P+Q=R+S$ make the following replacements

\begin{eqnarray}
p^{'}\rightarrow  {t\partial \over \partial p}, \qquad
q^{'}\rightarrow  {t\partial \over \partial q},\qquad
r^{'}\rightarrow  {\partial \over \partial r}\, ,\nonumber \\ 
\nonumber\\
s^{'}\rightarrow  {\partial \over \partial s},\qquad
u^{'}\rightarrow  {t\partial \over \partial u},\qquad
v^{'}\rightarrow  {t\partial \over \partial v}\, .
\end{eqnarray}
in this expression and evaluate the derivatives at
$p=q=r=s=u=v=0$.  This gives us 

\begin{eqnarray}
{\cal G}(D(A))& = & \sum_{{\tiny {PQRSUV;\,\,P'Q'R'S'U'V'}}} \left
(\frac{(U+2I+1)!(V+2I+1)!} {P! Q! R! S! U! V! (2I +1)})\times
 (\frac{P'! Q'! R'! S'! U'! V'! (2I' + 1)}{(U'+2I'+1)!(V'+2I'+1)!}\right )^{(1/2)}
\nonumber \\
&&\nonumber \\
&&\times (\frac{(V'+P'+Q'+1)!}{P'!Q'!R'!S'!U'!V'!(P'+Q'+1)})
\times D_{PQRSUV,\,\,P'Q'R'S'U'V'}^{(M=R+U,\,\,N=S+V)}(A) \times P!Q!R!S!U!V!\nonumber \\
\nonumber\\
&&\times \delta_{P^{'}P}\delta_{Q^{'}Q}\delta_{R^{'}R}\delta_{S^{'}S}\delta_{U^{'}U}\delta_{V^{'}V}\, ,\nonumber\\
\\
\label{FEQ1} 
&&=\sum_{M,\, N} \frac{(V+P+Q+1)!}{(P+Q+1)} t^{M+N} \chi^{(M,N)}_{SU(3)} (D(A))\, .\nonumber\\
\label{chi}
\end{eqnarray}
where we have used Eq.(\ref{GZL}) in writing Eq.(\ref{GZL}).

We next take up the expression on the r.h.s of Eq.(\ref{GFWD1}).
To exhibit the similarity of our starting point for the
computation of the characters of $SU(3)$ with that of
Schwinger's computation for $SU(2)$ we write this as

\begin{eqnarray}
{\cal G}(D(A))&=& (1 - V^{'}A^\dagger V)^{-2} {\exp} \left [ \frac{(R^{'} AR +
S^{'} A^\dagger S - uu'vv')}{(1 - V^{'}A^\dagger V)}\right ]\, .
\label{GFWD3}
\end{eqnarray}
where $A\in SU(3)$ and
\begin{eqnarray}
V&=&\pmatrix{s_q & s_p & v}\, ,\quad R=\pmatrix{r_p & r_q & u}\, ,\quad 
S=\pmatrix{s_q & s_p & v}\nonumber\\
\nonumber \\
V^{'}&=&\pmatrix{0 \cr 0 \cr v^{'}}\, ,\quad R^{'}=\pmatrix{r^{'}_p \cr r^{'}_q \cr u^{'}}\, ,\quad
S^{'}=\pmatrix{s^{'}_q \cr s^{'}_p \cr 0}\, .\nonumber \\
\end{eqnarray}

We now choose $A$ to be diagonal.  This is possible since every
unitary matrix can be diagonalized.  So let

\begin{eqnarray}
A=\pmatrix{e^{-i\theta{11}} & 0 & \cr 0 & e^{-i\theta_{22}} & 0\cr
0 & 0 & e^{-i\theta_{33}}}\, ,\nonumber\\
\\
\end{eqnarray}
where 
\begin{eqnarray}
\theta_{11}+\theta_{22}+\theta_{33} =0\, .
\end{eqnarray}

With this choice for the matrix $A$ the Eq.(\ref{GFWD3}) becomes

\begin{eqnarray}
{\cal G}(D(A))&=& \frac{1}{(1 - e^{i\theta_{33}}v'v)^{2}} {\exp} \left    [ 
\frac{(e^{-i\theta_{11}}{r_p}'r_p +e^{-i\theta_{22}}{r_q}'r_q 
+ e^{i\theta_{22}}{s_p}'s_p + e^{i\theta_{11}}{s_q}'s_q)
+u'(e^{-i\theta_{33}}u - uvv')}{(1 - e^{i\theta_{33}}v'v)}\right ]\, ,\nonumber \\
\nonumber \\
&=&\left (\frac{\exp  e^{-i\theta_{33}}u'u}{(1 - e^{i\theta_{33}}v'v)^{2}}\right )
\times {\exp} \left [ \frac{(e^{-i\theta_{11}}{r_p}'r_p 
+e^{-i\theta_{22}}{r_q}'r_q + e^{i\theta_{22}}{s_p}'s_p 
+ e^{i\theta_{11}}{s_q}'s_q)}{(1 - e^{i\theta_{33}}v'v)}\right ]\, ,\nonumber \\
\nonumber \\
&=&\left (\frac{\exp  e^{-i\theta_{33}}u'u}{(1 - e^{i\theta_{33}}v'v)^{2}}\right )
\times {\exp} \left [ \frac{(rr^{'} + ss^{'}e^{-i\theta_{33}})(e^{-i\theta_{11}}
pp^{'} + e^{-i\theta_{22}}qq^{'})}{(1 - e^{i\theta_{33}}v'v)}\right ]\, .\nonumber \\
\label{GFWD4}
\end{eqnarray}

We now replace $u^{'}$ with $t({\partial \over \partial u})$ and
evaluate the derivatives at $u=0$.  Similarly we replace $p^{'}$
with $t({\partial\over \partial p})$ and evaluate at $p=0$.
With $\lambda$ standing for the fraction $\frac{(rr^{'} +
e^{-i\theta_{33}}ss^{'})}{(1-e^{i\theta_{33}}vv^{'})}$ these
manipulations give us

\begin{eqnarray}
{\cal G}(D(A))&=&\left (\frac{\sum^\infty_{U=0}t^Ue^{-i\theta_{33}U}}
{(1 - e^{i\theta_{33}}v'v)^{2}}\right )\times 
{\exp} \left [ \lambda (te^{-i\theta_{11}}{\partial p};p +
te^{-i\theta_{22}}{\partial q};q)\right ]\, ,\nonumber \\
\nonumber\\
&&=\left (\frac{\sum^\infty_{U=0}t^Ue^{-i\theta_{33}U}}
{(1 - e^{i\theta_{33}}v'v)^{2}}\right )\times 
\frac{1}{1-t\lambda e^{-i\theta_{11}}} \times \frac{1}{1-t\lambda
e^{-i\theta_{22}}}\, , \nonumber \\
\nonumber\\
&&=\left (\frac{\sum^\infty_{U=0}t^Ue^{-i\theta_{33}U}}
{(1 - e^{i\theta_{33}}v'v)^{2}}\right )\times \frac{1}{t\lambda
(e^{-i\theta_{11}} - e^{-i\theta_{22}})}
\times \left \{ \frac{1}{1-t\lambda e^{-i\theta_{11}}} - \frac{1}{1-t\lambda
e^{-i\theta_{22}}} \right \}\, ,\nonumber \\
\nonumber \\
&&=\left (\frac{\sum^\infty_{U=0}t^Ue^{-i\theta_{33}U}} {(1 -
e^{i\theta_{33}}v'v)^{2}}\right )\times \frac{1}{t\lambda
(e^{-i\theta_{11}} - e^{-i\theta_{22}})}\nonumber\\
\nonumber\\
&&\times \left \{
\sum^\infty_{P+Q=0}t^{P+Q}\lambda^{P+Q}(e^{-i(P+Q)\theta_{11}} -
e^{-i(P+Q)\theta_{22}})\right \}\, ,\nonumber\\
\nonumber\\
&&=\left (\frac{\sum^\infty_{U=0}t^Ue^{-i\theta_{33}U}} {(1 -
e^{i\theta_{33}}v'v)^{2}}\right )
\times \frac{1}{t\lambda e^{-\frac{i}{2}(\theta_{11} + \theta_{22})}
(e^{-\frac{i}{2}(\theta_{11} - \theta_{22})} 
- e^{+\frac{i}{2}(\theta_{11}-\theta_{22})})}\nonumber\\
\nonumber\\
&&\times \left \{
\sum^\infty_{2I=0}t^{2I}\lambda^{2I}e^{-\frac{i}{2}2I(\theta_{11}+\theta_{22})}
(e^{-\frac{i}{2}(2I)(\theta_{11}-\theta_{22})} 
- e^{+\frac{i}{2}(2I)(\theta_{11}-\theta_{22})})\right \}\, ,\nonumber \\
\nonumber\\
&&=\left (\frac{\sum^\infty_{U=0}t^Ue^{-i\theta_{33}U}} {(1 -
e^{i\theta_{33}}v'v)^{2}}\right ) \times \sum^\infty_{2I=0}t^{(2I-1)}\lambda^{(2I-1)}
e^{\frac{i}{2}(2I-1)\theta_{33}}
\times \frac{\sin \frac{2I}{2}(\theta_{11}
-\theta_{22})}{\sin \frac{1}{2}(\theta_{11}-\theta_{22})}\, ,\nonumber\\
\nonumber\\
&&=\left (\sum^\infty_{U=0}t^Ue^{-i\theta_{33}U}\right ) 
\times \sum^\infty_{2I=0}t^{(2I-1)}\frac{(rr^{'}+e^{-i\theta_{33}}ss^{'})^{(2I-1)}}
{(1-te^{i\theta{33}}vv^{'})^{(2I+1)}}
\times e^{\frac{i}{2}(2I-1)\theta_{33}}
\times \frac{\sin \frac{2I}{2}(\theta_{11}
-\theta_{22})}{\sin \frac{1}{2}(\theta_{11}-\theta_{22})}\, .\nonumber \\
\end{eqnarray}

We will now work on the remaining variables.  So replace $v^{'}$
by ${\partial \over \partial v}$, $r^{'}$ by ${\partial
\over \partial r}$ and $s^{'}$ by ${\partial \over \partial s}$
and evaluate the derivatives at $r=s=v=0$.  Doing this we get

\begin{eqnarray}
{\cal G}(D(A))&=&\left (\sum^\infty_{U=0} \sum^\infty_{2I=0} \sum^{\infty}_{V=0}
\sum_{R+S=2I} \frac{(2I+V)!}{(2I)}
t^{(2I+U+V-1)} e^{-i\theta_{33}(\frac{-(2I-1)}{2}+S+U-V)} \right )\nonumber\\
\nonumber\\
&&\times \frac{\sin \frac{2I}{2}(\theta_{11} - \theta_{22})}
{\sin \frac{1}{2}(\theta_{11}-\theta_{22})}\, ,\nonumber\\
\nonumber\\
&&=\left (\sum^\infty_{U=0} \sum^\infty_{2I+1=0} \sum^{\infty}_{V=0}
\sum_{R+S=2I} \frac{(2I+V+1)!}{(2I+1)}
t^{(M+N)} e^{-i\theta_{33}(\frac{-2I}{2}+S+U-V)} \right )\nonumber\\
\nonumber\\
&&\times \frac{\sin \frac{2I+1}{2}(\theta_{11} - \theta_{22})}
{\sin \frac{1}{2}(\theta_{11}-\theta_{22})}\, .
\label{ourchi-1}
\end{eqnarray}

Comparing the above equation Eq.(\ref{ourchi-1}) with
Eq.({\ref{chi}}) we deduce that

\begin{eqnarray}
\chi^{(M,N)}_{SU(3)} (D(A))&=&\sum_{R+S=2I} e^{-i\theta_{33}(\frac{-2I}{2}+S+U-V)} 
\times \frac{\sin \frac{2I+1}{2}(\theta_{11} - \theta_{22})}
{\sin \frac{1}{2}(\theta_{11}-\theta_{22})}\, ,\nonumber\\
\nonumber\\
&&=\sum^{M}_{R=0}\sum^{N}_{S=0} e^{-i\theta_{33}(\frac{-2I}{2}+S+U-V)} 
\times \frac{\sin \frac{2I+1}{2}(\theta_{11} - \theta_{22})}
{\sin \frac{1}{2}(\theta_{11}-\theta_{22})}\, ,\nonumber\\
\nonumber\\
&& = e^{i\theta_{33}(2m_{13}-m_{23})} 
\sum^{m_{13}}_{m^{'}_{12}=m_{13}-m_{23}} \sum^{m_{13}-m_{23}}_{m^{'}_{22}=0} 
e^{-3i\theta_{33}(\frac{m^{'}_{12}+m^{'}_{22}}{2})} 
\times \frac{\sin \frac{m^{'}_{12}-m^{'}_{22}+1}{2}(\theta_{11} - \theta_{22})}
{\sin \frac{1}{2}(\theta_{11}-\theta_{22})}\, .\nonumber\\
\nonumber\\
\label{ourchi}
\end{eqnarray}
where use has been made use of Eq(\ref{QML}) and
Eq.(\ref{our-GZ}) in arriving at Eq.(\ref{ourchi}).

The Eq.(\ref{ourchi}) can be recast in the form given by
Eq.(\ref{gmchi}) by the replacement

\begin{eqnarray}
m_{23}\rightarrow  m_{13}-m_{23}\,.
\end{eqnarray}

Thus

\begin{eqnarray}
\chi^{(m_{13}-m_{23}, m_{23}, 0)}_{SU(3)} (D(A)) = e^{i\theta_{33}(m_{13}+m_{23})} 
\sum^{m_{13}}_{m^{'}_{12}=m_{23}} \sum^{m_{23}}_{m^{'}_{22}=0} 
e^{-3i\theta_{33}(\frac{m^{'}_{12}+m^{'}_{22}}{2})} 
\times \frac{\sin \frac{m^{'}_{12}-m^{'}_{22}+1}{2}(\theta_{11} - \theta_{22})}
{\sin \frac{1}{2}(\theta_{11}-\theta_{22})}\, .\nonumber\\
\end{eqnarray}
where the representations of $SU(3)$ are given by 
\begin{eqnarray}
(p, q) = (m_{12}-m_{23}, m_{23})\, .
\end{eqnarray}

With this we complete our derivation of Weyl's character formula
for the group $SU(3)$.

\section{Discussion}
Our aim in this paper has been to compute the Weyl's formula for
the characters of $SU(3)$ by using a method similar to the one
Schwinger used in deriving the character formula of $SU(2)$.  As
was noted by us in the introduction, Weyl's formula has been
known for a quite long time now and it has been derived many
times over by many people by using a vareity of methods.  But to
our knowledge ours is the first derivation to have obtained
Weyl's formula from a generating function for the matrix
elements of $SU(3)$.  For this purpose we made use of our
modified scalar product Eq.(\ref{GIP}) rather than the one
(Eq.(\ref{GIP1})) we had used in our erlier computations\cite{JSPHSS}.
But we need to stress the point that we would get the same
results even if we had done computations using our erlier
(unmodified) scalar product.

Our generating function itself is part of a calculus for doing
computations on $SU(3)$.  Since this calculus can be extended to
higher groups also we expect the generating function method
described in this paper also to carry over to higher groups.  

We can try a slightly different approach in generalizing these
computations to higher groups.  We can start with Weyl's formula
for the general case and then by working back we may be able to
recover the two tools of our calculus for the general case.
These two tools are (i). a generating function for the basis
states and (ii). a scalar product for the basis states.  Since
Weyl computed his formula by using group-invariant integrations
and since his character formula for $SU(n)$ seems to hide in
itself, at least in the special cases of $SU(2)$ and $SU(3)$, a
generating function for the basis as well as a presription for a
scalar product we may be able to connectup our constraint
Eq.(\ref{z.w}) with his construction on the group manifold and
our 'auxiliary' measure with the group-invariant measure itself.
This way one may be able to tieup the concept of a model space
of a group with a suitable realization of the group manifold.

With regards to applications we just mention one possibilty
which may be worth trying.  We know that in computing some
partition functions one has to make use of group characters and
measures.  In the normal way of doing things the integrations
over the group variables cannot be done exactly.  But in our
case the characters as well as the ('auxiliary') measure can be
written in the form of exponential functions and since these can
be combined with the partition function exponential one may be
able to evaluate the partition function more exactly for all the
IRs of the group and then extract the information regarding the
IR we are actually inetrested in as we had done in the case of
computing Clebsch-Gordan coefficients\cite{JSPHSS}.

\renewcommand{\theequation}{A.\arabic{equation}}
\setcounter{equation}{0}

\newpage

{\large\bf Appendix : Weyl's Character Formula for $SU(2)$ - 
Schwinger's Derivation}

Below we reproduce Schwinger's derivation of the Weyl's
character formula of $SU(2)$ \cite{SJ} almost verbatim.

Let $u\in SU(2)$ then, for Schwinger the generating function for
Wigner's $D$-matrix elements of $SU(2)$ is given by

\begin{eqnarray}
\exp (x^*uy)\, .
\end{eqnarray}

Formally this can be written as 

\begin{eqnarray}
\exp (x^*uy) = \sum_{jm}\phi_{jm}(x^*)U^{(j)}_{mm^{'}}\phi_{jm^{'}}(y)\, ,
\label{xuy}
\end{eqnarray}
where $\phi{jm}(x)$ and $\phi{jm^{'}}$ are basis states of the
IR labelled by the quantum number $j$ and $U^{(j)}_{mm^{'}}$ is
the matrix element of $u$ in the IR labelled by the quantum
number $j$.  As is well known for a fixed value of $j$ the range
of $m$ is, $-j\leq m \leq +j$.

For simplicity we shall assume the reference system to be so
chosen that $u$ is a diagonal matrix, with eigenvalues $e^{\pm
(i/2\gamma )}$.  We replace $x^*_\zeta$ with
$t(\frac{\partial}{\partial y_\zeta})$ and evaluate the
derivatives at $y_\zeta = 0$.  According to 

\begin{eqnarray}
\phi_{jm}(\frac{\partial }{\partial y}\phi_{jm^{'}}(y)]_{y_\zeta
= 0} = \delta_{m\, , m^{'}}\, .
\end{eqnarray}

Substituting this in the above Eq.(\ref{xuy}) we get 

\begin{eqnarray}
\left. \sum_j t^{2j} \chi^{(j)} = \exp (te^{-(i/2)\gamma} {\partial
\over \partial y_1}\, ; y_1 ) \cdot \exp (te^{(i/2)\gamma}
{\partial \over \partial y_2 }\, ; y_1 )\right ]_{y_\zeta = 0}\, .
\end{eqnarray}
in which the notation reflects the necessity of placing the
derivatives to the left of the powers of $y_\zeta$.  Now 

\begin{eqnarray}
\exp \left ( \lambda {\partial \over \partial y }\, ; y \right )
=\sum^\infty_{n-0} \frac{\lambda^n }{n!} \left ( {\partial \over
\partial y} \right )^n y^n = \sum^\infty_{n=0} \lambda^n =
\frac{1}{1-\lambda }\, .
\end{eqnarray}
and therefore
\begin{eqnarray}
\sum_j t^{2j} \chi^{(j)} &=& \frac{1}{1-t.\exp \left (
\frac{-i}{2}\gamma \right )}\times \frac{1}{1-t 
.\exp \left (\frac{i}{2}\gamma \right )}\nonumber \\
\nonumber \\
&&=\frac{1}{1-2t\cdot \cos\frac{1}{2}\gamma + t^2}
\end{eqnarray}
which is a generating function for the $\chi^{(j)}$.  On writing 
\begin{eqnarray}
\frac{1}{1-t.\exp \left (\frac{-i}{2}\gamma \right )}
\times \frac{1}{1-t .\exp \left (\frac{i}{2}\gamma \right )}
 = \frac{1}{2it\cdot \sin \frac{1}{2}\gamma }
\left [ \frac{1}{1-t.\exp \left ( \frac{i}{2}\gamma \right )} - \frac{1}{1-t 
.\exp \left (\frac{-i}{2}\gamma \right )}\right ]\, .
\end{eqnarray}
and expanding in powers of $t$, one obtains
\begin{eqnarray}
\chi^{(j)}(\gamma )=\frac{\sin (j+\frac{1}{2})\gamma}{\sin
\frac{1}{2}\gamma }\, .
\end{eqnarray}

\end{document}